\documentclass[sigconf]{acmart}

\AtBeginDocument{%
  \providecommand\BibTeX{{%
    \normalfont B\kern-0.5em{\scshape i\kern-0.25em b}\kern-0.8em\TeX}}}

\setcopyright{acmcopyright}
\copyrightyear{2021}
\acmYear{2021}
\acmDOI{xx.xxxx/xxxxxxx.xxxxxxx}

\acmConference[DSSG '21: The 3rd Workshop on Data Science for Social Good]{DSSG '21: The 3rd Workshop on Data Science for Social Good}{August 14, 2021}{Virtual Event}
\acmBooktitle{DSSG '21: The 3rd Workshop on Data Science for Social Good, August 14, Virtual Event}
\acmPrice{15.00}
\acmISBN{978-x-xxxx-xxxx-x/YY/MM}

\usepackage{multirow}
\usepackage{caption}
\usepackage{subcaption}
\usepackage{comment}

\begin{document}

%%
%% The "title" command has an optional parameter,
%% allowing the author to define a "short title" to be used in page headers.
\title[Experiences with the Introduction of AI-based tools for Moderation Automation]{Experiences with the Introduction of AI-based Tools for Moderation Automation of Voice-based Participatory Media Forums}

%%
%% The "author" command and its associated commands are used to define
%% the authors and their affiliations.
%% Of note is the shared affiliation of the first two authors, and the
%% "authornote" and "authornotemark" commands
%% used to denote shared contribution to the research.
%%
%% By default, the full list of authors will be used in the page
%% headers. Often, this list is too long, and will overlap
%% other information printed in the page headers. This command allows
%% the author to define a more concise list
%% of authors' names for this purpose.

\author{Aman Khullar}
\email{aman.khullar@oniondev.com}
\affiliation{%
  \institution{Gram Vaani}
  \country{India}
}

\author{Paramita Panjal}
\email{paramita.panjal@gramvaani.org}
\affiliation{%
  \institution{Gram Vaani}
  \country{India}}

\author{Rachit Pandey}
\email{rachit.pandey@oniondev.com}
\affiliation{%
  \institution{Gram Vaani}
  \country{India}}
  
\author{Abhishek Burnwal}
\email{abhishekburnwal2@gmail.com}
\affiliation{%
  \institution{IIT Delhi}
  \country{India}}
  
\author{Prashit Raj}
\email{cs1170359@cse.iitd.ac.in}
\affiliation{%
  \institution{IIT Delhi}
  \country{India}}
  
\author{Ankit Akash Jha}
\email{ankit.jha@alumni.iitd.ac.in}
\affiliation{%
  \institution{IIT Delhi}
  \country{India}}
  
\author{Priyadarshi Hitesh}
\email{priyadarshi.hitesh.mcs19@cse.iitd.ac.in}
\affiliation{%
  \institution{IIT Delhi}
  \country{India}}
  
\author{R Jayanth Reddy}
\email{r.jayanth.reddy.cs116@cse.iitd.ac.in}
\affiliation{%
  \institution{IIT Delhi}
  \country{India}}
  
\author{Himanshu}
\email{himanshurewar14@gmail.com}
\affiliation{%
  \institution{IIT Delhi}
  \country{India}}
  
\author{Aaditeshwar Seth}
\email{aseth@cse.iitd.ac.in}
\affiliation{%
  \institution{Gram Vaani, IIT Delhi}
  \country{India}}

\renewcommand{\shortauthors}{Khullar et al.}

%%
%% The abstract is a short summary of the work to be presented in the
%% article.
\begin{abstract}
Voice-based discussion forums where users can record audio messages which are then published for other users to listen and comment, are often moderated to ensure that the published audios are of good quality, relevant, and adhere to editorial guidelines of the forum. There is room for the introduction of AI-based tools in the moderation process, such as to identify and filter out blank or noisy audios, use speech recognition to transcribe the voice messages in text, and use natural language processing techniques to extract relevant metadata from the audio transcripts. We design such tools and deploy them within a social enterprise working in India that runs several voice-based discussion forums. We present our findings in terms of the time and cost-savings made through the introduction of these tools, and describe the feedback of the moderators towards the acceptability of AI-based automation in their workflow. Our work forms a case-study in the use of AI for automation of several routine tasks, and can be especially relevant for other researchers and practitioners involved with the use of voice-based technologies in developing regions of the world. 
% With advancements in speech recognition and natural language processing, interactive voice response systems have became a reality. However the rural population in the developing countries aren't well equipped or literate enough to use the existing systems. Therefore a need arises to run interactive information services for less-literate populations in rural regions. In this context, we describe our experience to automate a pre-existing interactive information service platform named GramVaani. We trained a system on data acquired by the GramVaani team and marked by human moderators. Our ultimate goal is now to use this data to automate the existing answer retrieval system so that the questions can be answered in real time by retrieving an appropriate answer from the corpus of questions and answers available so far. The first part in doing so is to automate the moderation process i.e. which is labelling the data based on it's quality, gender, theme and location. This will substantially increase the efficiency of the moderators and reduce human intervention. In this paper we will focus on the primary moderation process based on quality of the audio.
\end{abstract}

%%
%% Keywords. The author(s) should pick words that accurately describe
%% the work being presented. Separate the keywords with commas.

%% A "teaser" image appears between the author and affiliation
%% information and the body of the document, and typically spans the
%% page.
%%
%% This command processes the author and affiliation and title
%% information and builds the first part of the formatted document.
\settopmatter{printacmref=false}
% \setcopyright{none}
% \renewcommand\footnotetextcopyrightpermission[1]{}
% \pagestyle{plain}
\maketitle

\section{Introduction}
Voice-based discussion forums using IVR (Interactive Voice Response) systems have emerged as a useful modality for information sharing among low-income and less-literate populations in developing regions \cite{patel2010avaaj, vashistha2012ivr, raza2012viral, moitra2016design}. These forums are accessible through ordinary phones, not smartphones, and do not require the internet. Users can call and listen to audio messages, and respond by recording their own message which can be published back on the forum. Voice forums have found wide application in citizen journalism \cite{mudliar2012emergent, moitra2016design}, social accountability and grievance redressal \cite{marathe2015revisiting, chakraborty2017findings}, agricultural and livelihood information \cite{patel2010avaaj}, behavior change communication \cite{chakraborty2019experiences}, and cultural expression \cite{koradia2013gurgaon}, among others. 

Content moderation is a crucial element in the operation of voice forums. Most projects cited above employ a team of content moderators who listen to the recorded messages, and take action. They may reject the audio for poor audio quality or empty recordings, lightly edit the audio for publication on the platform, evaluate against editorial guidelines followed by the platform managers on aspects such as hate-speech or misinformation, annotate the recordings with tags or transcriptions, etc. Some of these steps are similar to content moderation done on internet-based participatory media platforms, where a combination of algorithmic and manual checks are used to filter inappropriate messages \cite{facebook_content, gorwa2020algorithmic}. Depending upon the platform design, peer users may be involved in a prior step to flag messages for further inspection \cite{crawford2016flag}, reasons behind the moderator actions may or may not be transparently disclosed to the users \cite{crawford2016can}, and users may be offered guidance by the moderators to submit more acceptable messages \cite{moitra2016design}.

%Internet-based participatory media platforms have led to several societal benefits which have empowered underserved communities, by allowing them to voice their opinions, concerns and grievances and help bridge prevailing inequities. The backbone of these platforms is the content that is promulgated through them --- the extent of influence this content can have on its users justifies the need for content moderation.

%Content moderation is the process of using formal editorial policies to ensure politically motivated content, misinformation, incomprehensible content, hate speech, incivil dialogue and other socially detrimental content is not published on a public platform. This task is usually carried out by a team of trained content moderators who read, watch or listen to the incoming content and then accept or reject the content based on their guiding policies. In some platforms, the moderators maintain accountability of their decisions by explaining the reason for rejection to the users and hence suggesting better contribution practices to the user.

We explore the feasibility of automation of some parts of the moderation process on voice-based discussion forums. We build machine learning based models to detect and reject empty audio recordings, determine the gender of the speaker, then use commercial ASR (Automatic Speech Recognition) APIs to obtain the transcript of the audio, and extract metadata from the transcript such as locations mentioned in the audio. In this study, we do not examine research topics on detection of hate-speech or fake news. We examine whether automation of the content operations listed above, can improve the efficiency of the content moderators so that they can devote more time to complex editorial checks rather than some of the above mundane tasks. Our evaluation is done in the context of several voice-based discussion forums operated by the social enterprise Gram Vaani. Gram Vaani employs about fifteen full-time content moderators who among other tasks manage content moderation on 40+ discussion forums that cumulatively generate approximately 1000 voice recordings each day. Through our evaluation, we examine whether automation of some of the tasks was found to be acceptable and useful by the content moderators, whether the moderation process can be entirely automated, and changes that the automation tools brought about in the moderation practices. 

%As any participatory media platform grows and amasses more users, the work-load on the moderators increases significantly and the advancements in vision, speech and text technologies and their appropriate use in content moderation offers some respite. This study describes our experiences of building machine learning based classifiers to facilitate content moderation of audio recordings and the lessons we learned by our deployment of these classifiers at X --- a voice based participatory IVR platform.

%X is an IVR-based interactive participatory media platform. It is free for users and works on ordinary feature-phones. The users give a missed call on a number and receive a call-back. Through this call they are able to listen to a mix of ‘User Generated Content (UGC)’, contributed by other users, as well the in-house generated, ‘Studio Generated Content (SGC)’. The users also have an option to record their own voice which is then passed on to the moderators who listen to the content and publish or reject the content based on their editorial policy.

%In order to facilitate the work of the moderators moderate incoming user generated content on MV, we deployed machine learning classifiers to check for acceptable audio content quality (non-blank and non-noisy) and extract some metadata like gender, theme, location and speech transcription automatically, which leaves room for the moderators to spend time on building and exercising an editorial policy.

Our main contributions are about interesting insights we gained in building machine learning models, a careful understanding about the introduction of AI-based methods in the moderation process, and a detailed time-cost analysis to evaluate any financial or time-saving benefits with automation of the moderation process. We find that with the current accuracy of speech recognition systems, moderation automation can reduce the workload of moderators but it certainly cannot replace them. This niche area of work presents an interesting case-study about AI-based methods and humans complementing each other. We have also open-sourced our models and made them available as a python library for the benefit of other researchers and practitioners working with voice-based discussion forums.

\section{Related Work}
Content moderation is an integral part of voice-based participatory media platforms, with each platform having its own moderation policies to regulate permissible content on the platform. One such platform called Mobile Vaani \cite{MV}, in whose context our research has been conducted, has fifteen full-time content moderators who review the audio content through a web-based content management system \cite{seth2020reflections}. CGNet Swara \cite{mudliar2012emergent} is another IVR platform with full-time moderators who check content quality and relevance, in line with the platform goals. Similarly, Avaaj Otalo \cite{patel2010avaaj} and IVR Juction \cite{vashistha2012ivr} are voice forums which work for societal development and also have a set of dedicated moderators for content moderation.

As these platforms scale, a proportional increase in moderation resources is required. Increasing the size of the moderation team has its own limitations with management and funding constraints. Further, several tasks related to content moderation can be routine and mundane, drawing attention away from more crucial aspects that require human judgement. On the internet, community or distributed moderation options have been explored on platforms such as Reddit, Stack Overflow, and Slashdot. Several studies \cite{chandrasekharan2017you, chandrasekharan2018internet, lampe2004slash} have demonstrated the efficacy of these moderation policies and how they are in general a viable option for scaling of public forums. Sangeet Swara \cite{vashistha2015sangeet} and Gurgaon Idol \cite{koradia2013gurgaon} were among early voice-based forums to experiment with some aspects of community moderation as well, such as ranking of audio content based on its quality, and identifying the broad topic category of the content. 

Our work is with a similar goal to help scale voice-based forums. We however focus on the automation of routine tasks specific to voice-based discussion forums, such as the identification of empty or noisy audio recordings, and metadata extraction from automatically transcribed speech recordings. We do not explore content-based moderation for aspects like hate-speech or misinformation, since this requires human judgement and is specifically why we want to free up time for the moderators to be able to focus on these aspects, as well as explore distributed moderation in the future where community volunteers can also participate in making these judgements. Our work is closer to platforms like BSpeak, ReCall and Respeak \cite{vashistha2018bspeak, vashistha2019recall, vashistha2017respeak}, which also used automatic speech recognition to achieve a reduced transcription cost and faster turnaround time for data collection tasks allocated to visually-impaired people. We evaluate the benefits that can be gained from recent advances in speech recognition and natural language processing for low-resource languages while acknowledging the risks of full automation in decision support systems \cite{de2020case}. For tasks like automatically rejecting blank or noisy audio, and gender identification, we build upon techniques in audio classification using signal processing \cite{lu2002content, guo2003content} and convolutional neural networks \cite{hershey2017cnn, lee2009unsupervised}. For metadata extraction tasks, we operate on transcripts obtained through commercial speech recognition APIs and build topic classification and named entity recognition methods \cite{wang2012baselines, nadeau2007survey}. We deployed these tools in the Gram Vaani \cite{GV} technology stack and ran several experiments and feedback studies to understand their impact on the moderation workflow.

% which suggest why we also carry out topic classification and location entity recognition as part of moderation automation ML models stack --- though we have trained all the models using our custom dataset for higher accuracy. In our work, we also evaluate the benefits from using off-the-shelf automatic speech recognition technology and its impact in the day-to-day moderation activities.

% but instead of taking the path of utilizing community support, we operate in the context of a centralised moderation team. This makes it easier for us to conduct controlled studies, and the techniques can possibly be opened up for community or distributed moderation in the future. 

%We next present our key insights from the use of recent automation tools to support the field of centralised content moderation and also present a comprehensive evaluation of these tools to guide the future research in this field.

\section{Machine Learning Models}
% This platform receives an average of 10,000 calls daily (these are sometimes accompanied by outbound calls) and 1,000 daily voice messages. These messages are then moderated by a team of trained content moderators who listen to these recordings and publish on the MV platform if the audio is of suitable quality, does not contain any objectionable or false information and is pertinent to the audience being served by that platform. Each moderator listens to around 100-120 audio messages every day and 30\% of total incoming messages go on to be published on the MV platform. \par

% Based on the trenchant analysis of the audio message and the editorial policies, various metadata fields are added to the audio message. Some of the fields include - message decision (acceptance or rejection), speaker’s gender, speaker’s location, message topic, message rating (0 being the lowest and 5 being the highest), message transcription and message title. \par

After several years of running voice-based discussion forums that were moderated manually, Gram Vaani has accumulated a large annotated dataset of audio recordings. We henceforth refer to these as \emph{audio items} or simply \emph{items}. The dataset has been annotated with the following metadata:
\begin{itemize}
    \item Item state: Whether the audio recording was accepted for publication on the IVR or rejected. For most items, a reason for rejection is also marked of whether it was rejected because it was an empty audio recording, or it had too much background noise, or the recording was not articulate enough, or it was rejected for editorial reasons such as cyber bullying or hateful content.
    \item Gender: The gender of the speaker - Female, Male, Third gender, or a Group of speakers, identified by the moderators based on the voice of the speaker and any identification information they provided about themselves in the audio recording. 
    \item Location: The State, District, Sub-district and the village of the speaker, based on the details provided by the speaker in their recording.
    \item Tags: Tags to identify a broad topic are marked by the moderators, out of a large set of approximately 150 standard tags that have emerged over several years at Gram Vaani. Short-lived event specific tags are also developed. These tags are helpful for project staff and researchers to search for items.
    \item Rating: A qualitative assessment of the quality of the item, with a value between 1 to 5. The quality largely depends on the content spoken in the recording, of whether it provides some interesting or novel information, or a new viewpoint. 
    \item Transcription: This contains the transcription given by the moderators to the item, and may either be a full transcript or a short gist of what the user is speaking about. Moderation policies varying from project to project are used by the moderators to determine whether to give a gist or full transcript to the audio.
    \item Title: Each audio recording is also annotated with a title or headline given by the moderators.
%    \item Source: This field helps keep a track if the audio recording had been generated in-house or if it was a user contribution.
\end{itemize}

We use a subset of this data to build and evaluate machine learning models, as described next. 

%To build and test each of our moderation automation models, we use a subset of the complete data and use only the relevant metadata required to build that model. In all of our experiments, we ensure that there are no duplicate entries in our dataset to avoid double counting.

% The platform has received X calls since its inception in 2011 with its user base spread across various states in India, speaking in multiple languages. The most commonly received messages are in Hindi, Tamil and Kannada, prevalent in Central India and South India respectively. For our purpose, we have concentrated on the data received from Hindi speaking states and in order to incorporate the varying dialects, we have considered data from 4 states --- Bihar, Jharkhand, Madhya Pradesh and Uttar Pradesh while also separately considering the Nalanda district within Bihar.

% - A pie chart for rejection reasons?

\subsection{Blank Classifier}
\begin{table}[t]
  \caption{Confusion matrix for the blank classifier}
  \label{tab:blankcm}
  \begin{tabular}{ccl}
    \toprule
    Predicted Class $\rightarrow$ & Blank & Accepted\\
 True Class $\downarrow$ & &\\
 \midrule
 Blank & 632 & 19 \\
 Accepted & 33 & 2849 \\
  \bottomrule
\end{tabular}
\end{table}
% \subsection{Dataset}
The blank classifier attempts to classify whether a recording is empty or it has human speech. We used a subset of 17,000 audio recordings to build the train set, and 3,500 audio recordings to build the test set. We labeled an item as \emph{accept} if its publication state was an accept and it had a rating of 4 or 5. Items were labeled as \emph{reject} if their publication state was a reject and the reason for rejection was specified as blank. We ensured that we sampled these training and test-set audio recordings for both male and female speakers, and from across different geographic regions where Gram Vaani has been working, to provide diversity. 

We build a feature set for an audio as follows. We first segment the given audio recording into chunks having a frame size of 50ms and a step size of 25ms, and then use the pyAudioAnalysis library \cite{giannakopoulos2015pyaudioanalysis} to extract a set of 34 audio features for each audio chunk. These include the mean amplitude, entropy of the amplitude, spectral spread, MFCCs and other signal processing features. We then find the distribution of these features across all the chunks in an audio, and use the 4 quartiles for each feature as the actual features that are fed into a machine learning classifier. Using these features, we trained several classifiers and found the Random Forest Classifier followed by recursive feature elimination and hyper-parameter tuning, to provide an accuracy of 98.5\%, as shown in table \ref{tab:blankcm}, and a false negative rate of 1.15\%.

\begin{table}[t]
  \caption{Confusion matrix for the gender classifier}
  \label{tab:gendercm}
  \begin{tabular}{ccl}
    \toprule
    Predicted Class $\rightarrow$ & Male & Female\\
 Marked Class $\downarrow$ & &\\
 \midrule
 Male & 552 & 68 \\
 Female & 36 & 579 \\
  \bottomrule
\end{tabular}
\end{table}

\subsection{Gender Classifier}
We use the same methodology as above to train a gender classifier as well, that classifies based on the audio whether the speaker is a male or female. The classifier is intended to be applied only on non-blank audio that may be recommended for publication. To build the training set, we used a subset of around 4,700 audio recordings, with a balanced split between the Male and the Female classes, sampled across the different geographies covered by Gram Vaani.

For feature extraction, we identified a subset of 136 features using recursive feature elimination from the feature-set distribution used in the blank classifier, which we then used to train our machine learning model. We finally trained an SVM classifier and fine-tuned the hyper-parameters to obtain a test set accuracy of 91.6\% on around 1,200 audio recordings, as shown in table \ref{tab:gendercm}.

\begin{comment}
\textcolor{green}{To build a training set, we again use only items with \emph{ratings} of 4 and 5, and we ensure a balanced split between the Male and Female classes. The same methodology for feature extraction is used as with the blank classifier, and finally an SVM classifier is selected for its higher accuracy. An accuracy of 95.9\% is obtained on the test data, as shown in table \ref{tab:gendercm}.}
\end{comment}

\subsection{Transcription and Location}
\begin{table}[b]
\caption{Accuracy of custom location entity extractor as compared to Google's Dialog Flow suite}
\label{tab:acc_comp_stts}
\begin{tabular}{clccc}
\toprule
Entity &
   &
  \begin{tabular}[c]{@{}c@{}}Bad STT \\ (\%)\end{tabular} &
  \begin{tabular}[c]{@{}c@{}}Accuracy of\\  DF with \\ good STT\end{tabular} &
  \begin{tabular}[c]{@{}c@{}}Accuracy of\\ custom module\\ with good STT\end{tabular} \\
\midrule
\multirow{2}{*}{Location} & State    & \multirow{2}{*}{28.71} & 51.81 & 87.78 \\
                          & District &                        & 51.15 & 75.57 \\
\bottomrule
\end{tabular}
\end{table}
We used the Google ASR (Automatic Speech Recognition) API \cite{STT} to obtain transcripts for the audio recordings. These STT (Speech to Text) transcripts were then analyzed to extract location entities. We experimented with inbuilt entity extraction tools provided by Google's Dialog Flow suite \cite{dialogflow}, but did not get a very high accuracy, likely due to the use of local names. We therefore built our own custom entity extraction tool. 

Using a multilingual text processing library called Polyglot \cite{polyglot} which uses parts of speech rules for entity extraction, we fist obtain candidate location entities identified by Polyglot's Named Entity Recognition (NER) module. We then run direct and approximate string matching for these entities against the list of all Indian states, districts, and sub-districts, according to the 2011 Indian census \cite{census}. We updated the census dataset with a few changes such as for the state of Andhra Pradesh which was split into two states, and similarly for a few districts that were split into smaller districts. We also added name variations such as \emph{East/West} Champaran to the Hindi equivalent of \emph{Purbi/Pashim} Champaran. In case a Hamming distance similarity of less than 10 (empirically determined threshold) was found between a location entity spotted by Polyglot and the census entries, we identified that entity as a valid location. We also backfilled information such that if a district name was matched then we auto-filled the name of the corresponding state, or if a sub-district was matched then we auto-filled the name of the district and state. 

%Finally, in case we find any similarity between the Polyglot's location entities with the Census dataset, we initiate a process of \emph{location backfill}, which helps us fill the less precise location in cases where the user mentions the more precise location. For example, if the user mentions only the district in their recording, then we backfill the state and if the user mentions their subdistrict location then we backfill both the district as well as the subdistrict. However, in India since different states sometimes have districts with the same name, it is possible that backfilling step might give us more than one location and in such cases we return all the possible location values and consider our answer to be accurate if any of the values matches against the ground truth entity.

%We then used direct and approximate string matching heuristics in which we calculated the hamming distance between the polyglot's output location entities against the list of locations available from the Census dataset. Here, we consider the polyglot's output similar to any location in the Census dataset only if their hamming distance is less than 10 (which was empirically found to be most suitable).

Table \ref{tab:acc_comp_stts} reports the accuracy at the state and district levels, computed on a ground-truth of 425 datapoints. The accuracy provided by Dialog Flow was much lesser than the state level accuracy of 87.78\% and district level accuracy of 75.57\% obtained through our custom methods. These accuracies are computed on STT transcripts which contain meaningful location information. We found that between 25 to 30\% of the STT transcripts were empty or with incorrect information due to ASR inaccuracies.

\section{Time-Cost Analysis}
We next report our findings from deploying the models described in the previous section. We analyze the impact of introducing these AI-based methods on potential time and cost savings, and how the moderators reacted to incorporating the tools in their workflow. 

To build a study design, we first spoke to a few moderators to understand their daily workflow in detail. We found that moderators essentially perform four tasks. The first task involves a quick determination of whether an audio item may be publishable or not, based on its audio quality. Second, the moderators mark various metadata fields (such as the gender of the speaker, location, topic, title, and assign a qualitative rating between 1-5 based on the relevance of the item) associated with the item. To carry out this task, the moderators first listen to the audio and then mark the metadata. Third, the moderators listen to the item again to write its transcript. Here, the moderators determine whether they should just write a short gist or not even that, or should they give a full word-to-word transcription for the item. The moderators follow internal policies, such as to give a detailed transcript to interesting or unique items that may merit being shared with the wider project team, or for grievances related to government schemes which are often shared through social media channels to draw the attention of government administrators to the issue \cite{chakraborty2017findings}. Finally, the moderators may lightly edit the item to normalize its volume or remove periods of silences before publishing it back on the the IVR platform. 

%We conducted an ethnographic study before starting this experiment to understand how to effectively evaluate the time-saving offered by the machine learning models. 

In our study, we evaluate the impact of automation on the first three tasks only: namely, to identify and reject poor quality audio items, provide metadata annotation automatically to a few fields, and assist the moderators with providing an STT (Speech to Text) transcription obtained through the Google ASR APIs. We conduct this study in two rounds, with and without the automated tools being in place. In each round, we isolate the impact along two axes. First, we isolate the impact due to the three tasks of blank audio detection, metadata annotation, and transcription. It was not feasible to further breakdown the metadata annotation to different sub-parts because the moderators mark these fields in one go making it difficult to time their effort on each one separately. Second, for the transcription component, we separately analyze the impact depending upon whether the moderators gave a gist transcription or a full transcription. 

%We also break down our study into three parts with each evaluating the effect of automation with a different transcription policy. In each of the thee parts, we observe the difference in moderation time with/without automation and the difference in transcription time with/without automation. We tried breaking down the moderation time to just observe the effects of the gender classifier and the location entity extractor but through our initial ethnographic study we found that the Moderators mark several metadata fields simultaneously and it was not possible for them to report time taken to mark each individual field.

The study was carried out with the help of 13 moderators, with at least four moderators participating in the experiment each day, for a span of 33 days. The activity was carried out for 6 hours each day in an instrumented manner. The moderators were asked to use a stopwatch to note down the time taken in carrying out each of these tasks for each item that they moderated. They also noted the item-id (a numeric six digit unique identifier for each item) so that we could link their notes with other item details such as its audio duration and the accept or reject action taken upon the item. The moderators counted the time they take to listen to the audio as part of the overall task duration. We next present the results to understand the impact of the blank classifier in reducing the number of items put up to the moderators for review, the impact of the gender classifier and location entity extraction on the time spent in metadata annotation, and time savings made by providing an STT transcription. The study was carried out during a time when COVID-19 related topics had subsided in India and regular project activities had resumed, with no significant difference in the two study rounds of with and without automation. 

%\textcolor{red}{Moreover, the study was done during a steady state (in terms of the number of audio recordings being received on the IVR platform) with not much difference between the with/without automation phases. But the automation was most appreciated during the COVID-19 lockdown and described in more detail in section \ref{}}

\subsection{Blank classifier}
We first present the time-saving offered by the blank classifier. Out of 498 items received during the with-automation phase, the moderators rejected 174 (34.94\%) items. During the without-automation phase, the moderators received 712 items and rejected 437 (61.38\%) items. 

\begin{figure}[t]
  \centering
  \includegraphics[width=8cm]{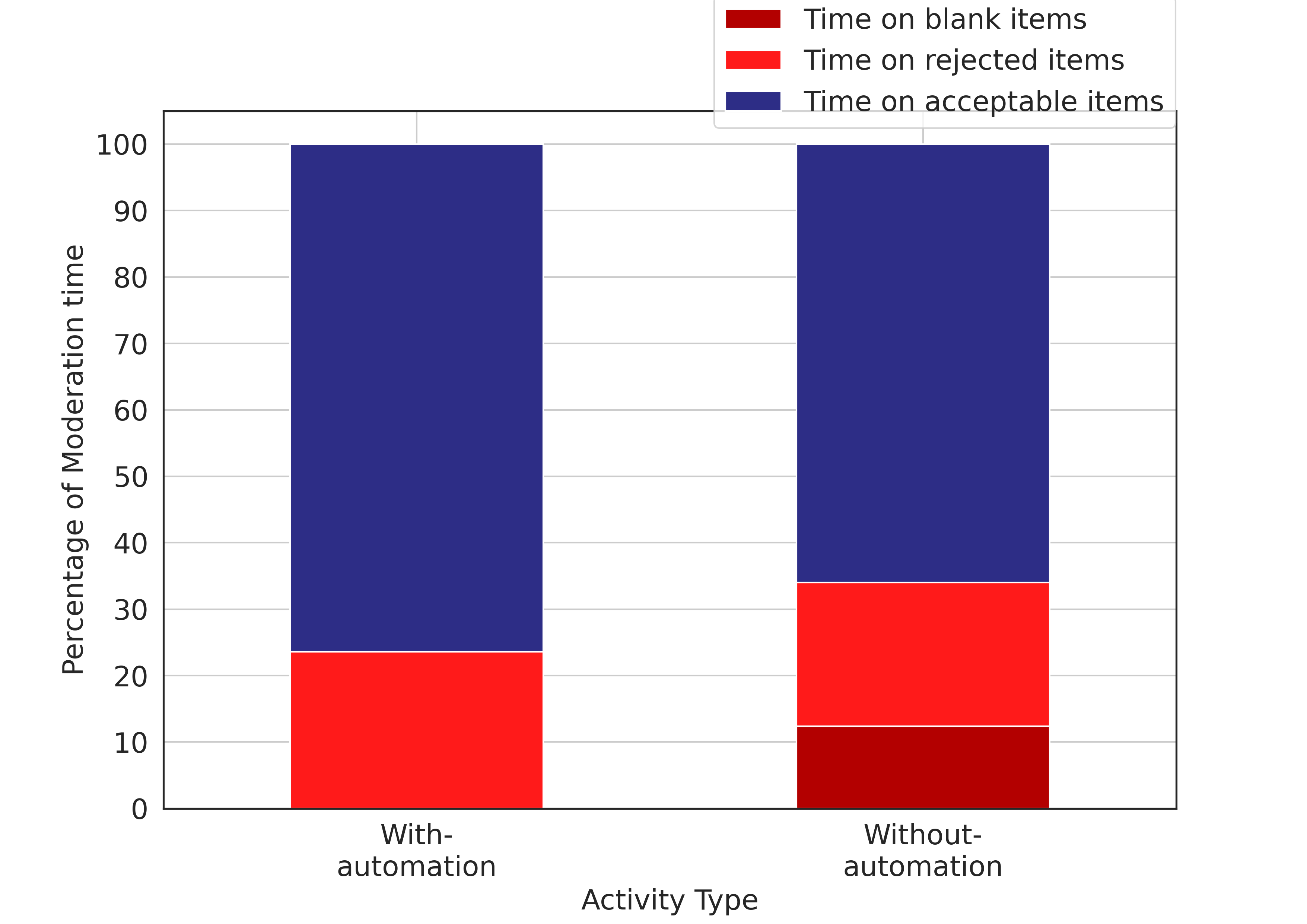}
  \caption{Overall time-saving offered by the Blank classifier}
  \label{fig:blank_gist_full}
  \Description{Time saving with the use of the blank classifier.}
\end{figure}

%\subsubsection{Blank Classifier}
Figure \ref{fig:blank_gist_full} shows the time-saved by the moderators: we found that the moderators spent 13\% less time in sifting through rejected items when the blank classifier was deployed, which provided a saving of 11.74 seconds per accepted item and amounted to a time-saving of 17.07\% in rejecting a blank item per accepted item. While the savings are modest, upon speaking with the moderators they reported that earlier they would do a quick check by waiting for the audio to be downloaded and then look at its waveform visualization to determine if it was blank or not, but with the automation they are able to save on the download wait time and move faster. Another interesting benefit they reported was that seeing less items for moderation helps reduce their anxiety. Especially during the COVID-19 lockdown in India, the IVR platforms were very heavily used, and a moderator reported:

``\emph{During the early days of Coronavirus, there used to be 500 items each day and we got anxious seeing so many items to moderate, but after ML (machine learning) was deployed, we see much lesser number of items. During Covid, ML helped us very much.}'' --- Senior Moderator, Gram Vaani, Gurgaon.

\subsection{Metadata annotation}

\begin{figure}[t]
  \centering
  \includegraphics[width=8cm]{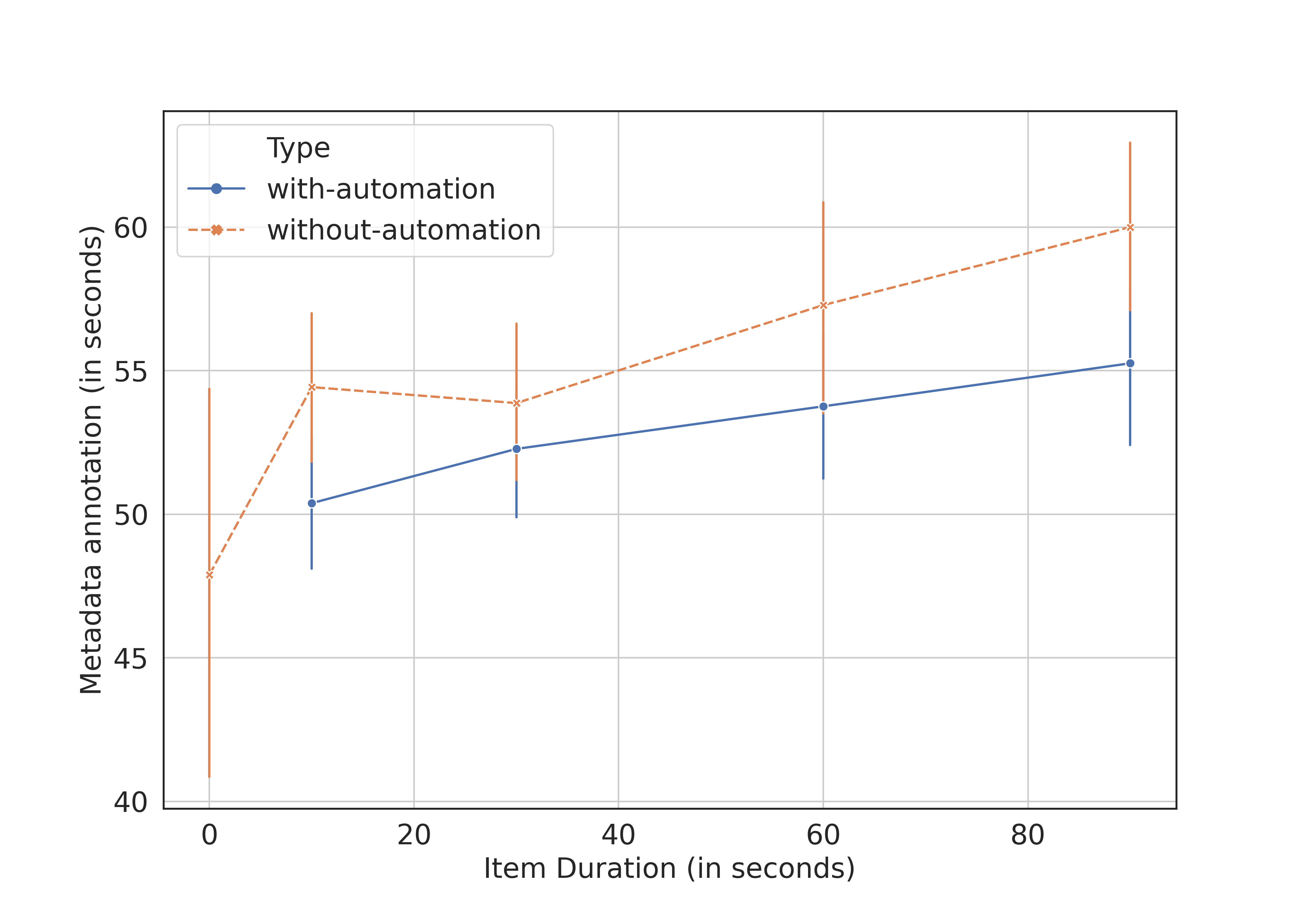}
  \caption{Time-saving offered by metadata annotation}
  \label{fig:metadata_line}
  \Description{Time saving with the use of metadata annotation.}
\end{figure}

Figure \ref{fig:metadata_line} shows the time-saving with metadata annotation deployed for gender marking and location entity extraction. In aggregate, moderators saved 5.97\% time with automation, amounting to an average of 3.34 seconds saved per item out of 56.04 seconds taken on average for metadata annotation per item.

\subsection{Transcription}

\begin{table}[b]
  \caption{Duration-wise time-saving in gist transcription}
  \label{tab:trans_gist_perc}
  \begin{tabular}{ccc}
    \toprule
    Item Duration (sec)& Avg. time saved (s) & Avg. time taken (s)\\
    \midrule
    10-20 & -1.25 & 59.56\\
    20-40 & 3.13 & 79.60\\
    40-60 & 11.52 & 81.80\\
    60-100 & 9.29 & 93.61\\
    Greater than 100 & 10.61 & 85.04\\
    
  \bottomrule
\end{tabular}
\end{table}

\begin{figure}[t]
  \centering
  \includegraphics[width=8cm]{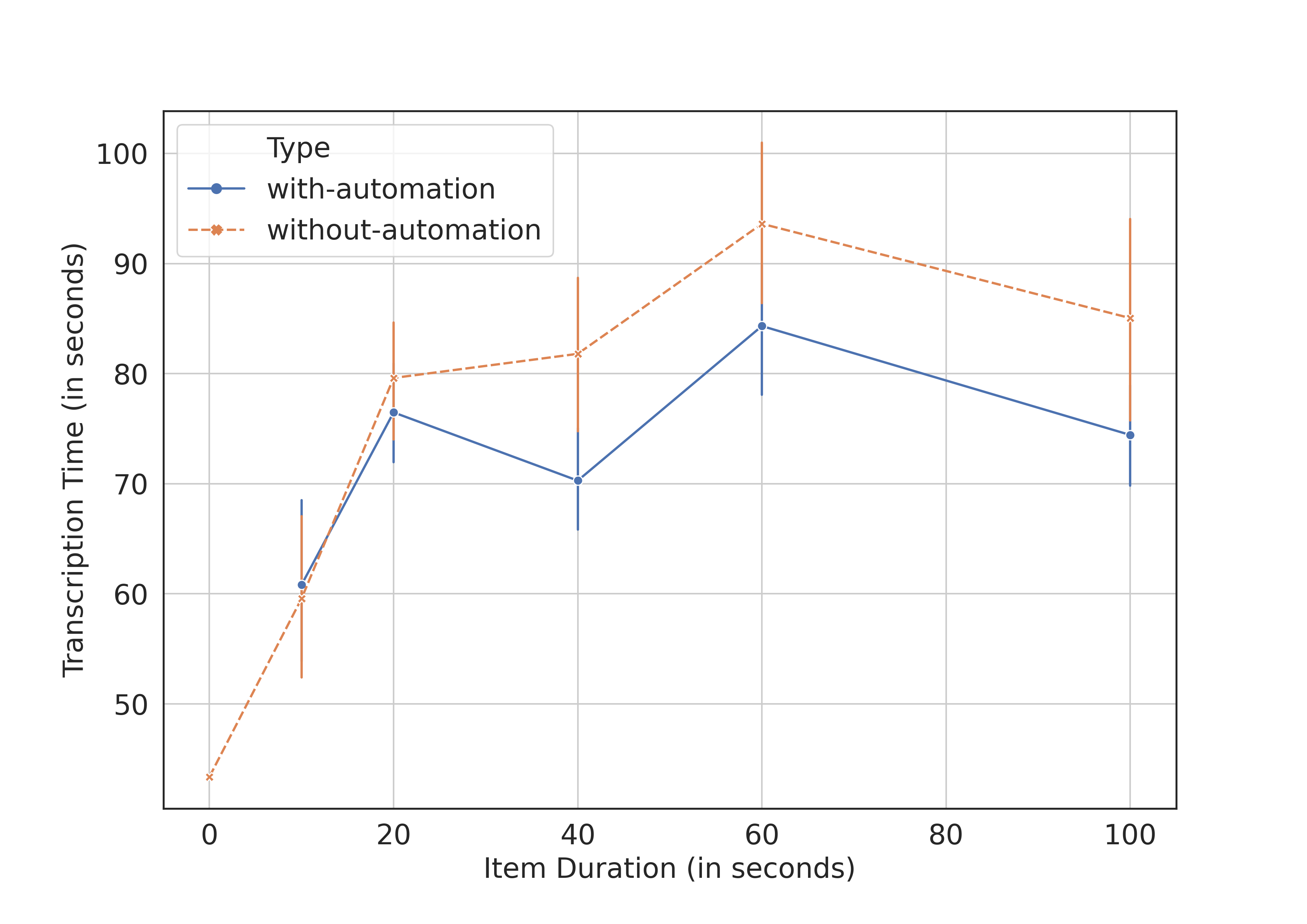}
  \caption{Time-saving offered by automated STT transcripts to write human-curated gist transcripts}
  \label{fig:trans_gist}
  \Description{Time saving due to use of automated STT transcripts in writing gist transcripts.}
\end{figure}

\subsubsection{Gist Transcription}
%This study was done in a span of 14 days in 2 parts, 7 days with-automation and 7 days without. The moderators gave gist transcriptions to 302 items (out of which 186 were accepted for publishing on the IVR) during the with-automation experiment, and a total of 350 recordings (out of which 139 were accepted) during the without-automation experiment. 

% \begin{figure}[b]
%   \centering
%   \includegraphics[width=\linewidth]{line_plot_moderation_jist_with_err_bars_paper.png}
%   \caption{Time-saving offered with the gender classifier and location entity module from gist transcription experiment}
%   \label{fig:gen_loc_gist}
%   \Description{Time saving because of the gender classifier and location entity module in with/without automation comparative study.}
% \end{figure}

%\subsubsection{Transcription time-saving}
The moderators gave gist transcriptions to 186 items during the with-automation experiment, and to 139 items during the without-automation experiment. To understand variation with the duration of audio items, we create broad bins for the recordings and analyze the time saving within each bin for when the STT transcripts are available to the moderators and when they are not. Figure \ref{fig:trans_gist} and table \ref{tab:trans_gist_perc} show the time savings which seem to become more significant for longer audio recordings. In aggregate, moderators saved 8.6\% time with automation, amounting to an average of 6.99 seconds saved per item out of 81.61 seconds taken on average for writing a gist transcription per item. The savings are modest though, and the moderators confirmed upon further discussion that the STT is useful in helping grab the name of the person or the location, but they prefer relying on well-practiced templates for writing gists. For example, if somebody records a song or poem, or appreciates a programme they have heard on Mobile Vaani, the moderators just write a line of the form: ``\emph{[name] from [location] \{talks about, sings, appreciates\} [topic]}''. Since the moderators anyway have to spend time in listening to the item, the STT does not offer substantial benefit in informing the moderators about the topic or offering text that can be copy-pasted to build the gist transcript. The STT is more useful if the moderators want to provide a full transcript, as described next. 

%And this benefit becomes more significant when the duration of the item increases as shown in table \ref{tab:trans_gist_perc}.

%One question that we did ask here that was that even though we found a time-reduction but the reduction is not of the order we would expect it to be since the transcription is available to the Moderators and in order to summarise it, all one expects them to do is copy and paste a part of the transcription. Upon asking this question to the Moderation team, we found out that in fact their transcription policy asks them to write a transcription in third person so they are able to benefit from the STT only with getting some parts of information like person's name, location and if they need to quote the speaker. Otherwise, they write the complete transcription in their own words, especially when the item is of shorter duration follows a common type of response heard on the platform for which they have a template response (for example when a person is praising about the content being published on the platform). In these cases, the moderators prefer to write the transcription completely rather than copying the STT and editing which they find more time-consuming.

\subsubsection{Full Transcription}

\begin{table}[b]
  \caption{Duration-wise time-saving in full transcription}
  \label{tab:trans_full_perc}
  \begin{tabular}{ccc}
    \toprule
    Item Duration (sec)& Avg. time saved (s) & Avg. time taken (s)\\
    \midrule
    10-20 & -0.57 & 135.26\\
    20-40 & -25.06 & 191.54\\
    40-60 & 39.16 & 316.11\\
    60-100 & 102.99 & 475.77\\
    Greater than 100 & 208.00 & 608.40\\
    
  \bottomrule
\end{tabular}
\end{table}

\begin{figure}[t]
  \centering
  \includegraphics[width=8cm]{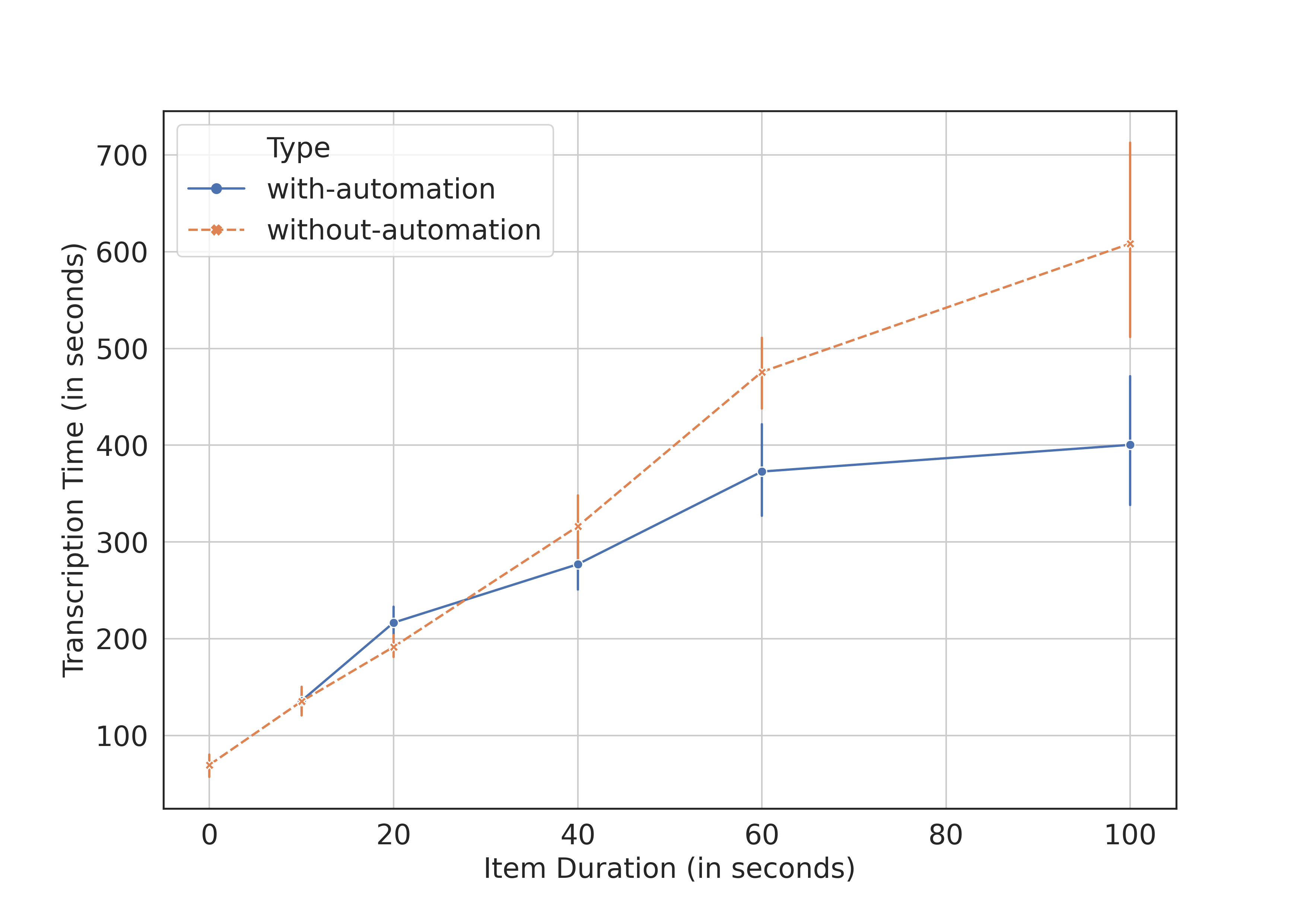}
  \caption{Time-saving offered by automated STT transcripts in writing human-curated full transcripts}
  \label{fig:trans_full}
  \Description{Time saving due to use of automated STT transcripts in writing full transcripts.}
\end{figure}
The moderators wrote full transcripts for 142 recordings during the with-automation experiment, and for 136 items during the without-automation experiment. Figure \ref{fig:trans_full} and table \ref{tab:trans_full_perc} shows the time-savings for audios of different duration. In aggregate, moderators saved 17.77\% time with automation, amounting to an average of 56.92 seconds saved per item out of 339.39 seconds taken on average for writing full transcription per item. These savings are much more than with the gist transcripts, as expected, but still quite modest. We were expecting that the moderators will be able to do a direct copy-paste of the STT transcripts but that was not the case. The moderators pointed out that the STT accuracy for some items could be very poor, while in other cases they did use the transcripts but had to correct it to fix the inaccuracies. The correction process itself is not straightforward, as one of the moderators explained: 

%An important finding here was that a direct copy-paste of the STT is not possible right now because of the inaccuracies in the STT which lead to erroneous/missing words in the STT. Moreover, we found that the problem is quite extensive and the Moderators prefer to write their own word-to-word transcription when the items are of shorter duration (which is also what we found, as shown in figure \ref{fig:trans_full}). As explained by one of the Moderators:

%were very much in line with what we expect the effect of speech-to-text transcription to be on writing word-to-word transcripts. Though we did find an increasing percentage in time-saving as the item duration increased (as shown in table \ref{tab:mod_full_perc}), they were again not of the order we were expecting them to be since this time it was a direct first person word-to-word transcription and ideally we were hoping to observe that all that Moderators would have to do is copy and paste the STT without even listening to the audio.

``\emph{We are not able to understand what the recording is talking about only with the help of ML and we need to listen to the audio and then pause the audio, remove the incorrect word, add the correct word and then play the audio again. All this takes a lot of time and we find it better to just sometimes write the word-to-word transcript ourselves}'' --- Senior Moderator, Gram Vaani, Gurgaon.

\begin{figure}[b]
  \centering
  \includegraphics[width=8cm]{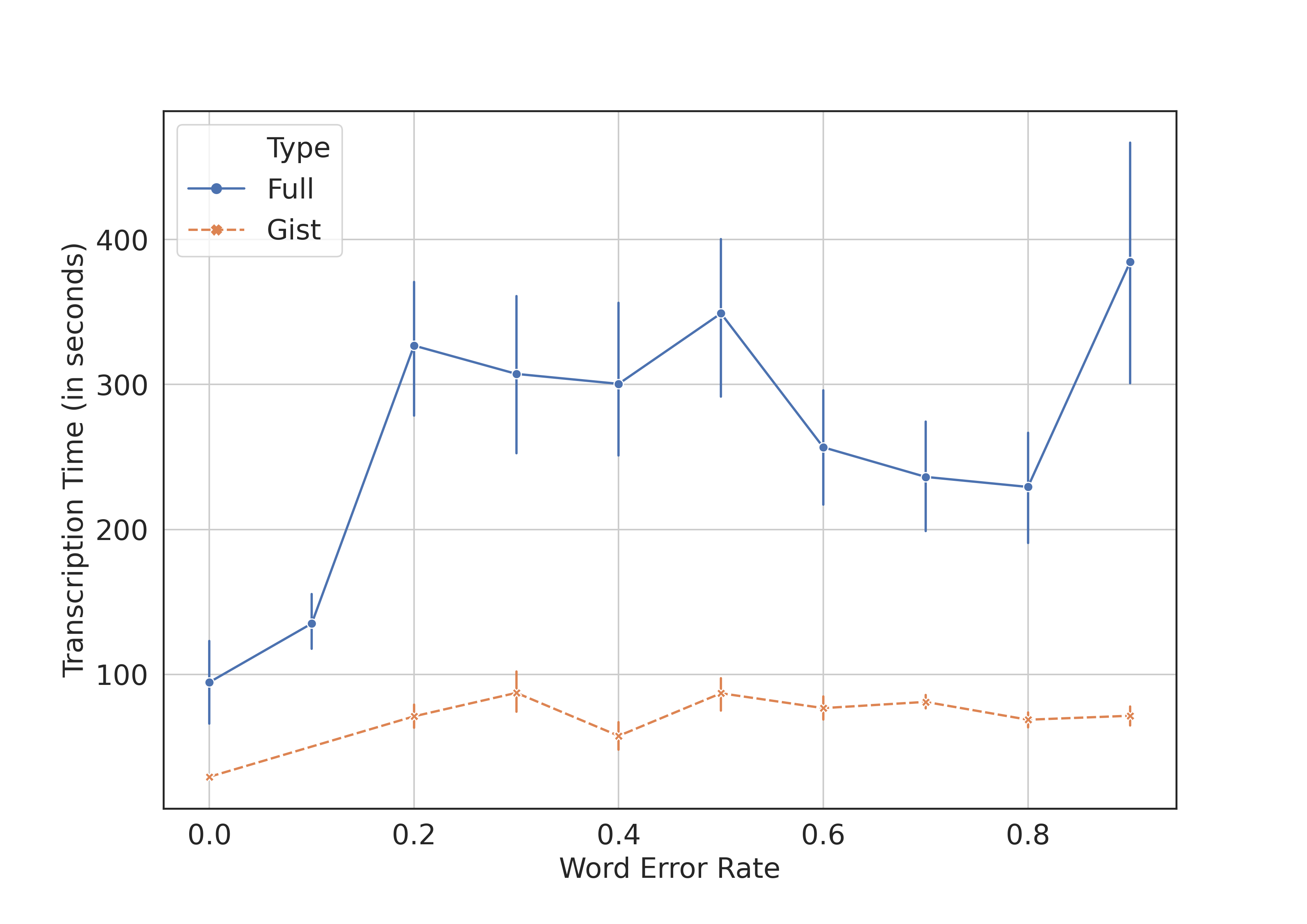}
  \caption{Affect of STT WER (Word Error Rate) on transcription time}
  \label{fig:wer}
  \Description{STT Word Error Rate vs. Gist and full transcription time}
\end{figure}

The moderators broadly stated that as long as word errors are in the range of 10-15\%, they find the STT to be useful, but otherwise they prefer writing the transcript from scratch and at best just copy parts of the STT. This also bore out in our analysis, shown in Figure \ref{fig:wer}, where we plot the WER (Word Error Rate) for the STT transcripts compared with a manual transcription. We can see that as the WER climbs above 10\%, the time taken by the moderators for writing a full transcript increases substantially, likely due to them resorting to transcribing the items themselves without relying on the STT transcripts. For gist transcripts on the other hand, since the moderators tend to write templatized transcripts, the variation with WER is not very significant.

% \begin{figure}[b]
%   \centering
%   \includegraphics[width=\linewidth]{ratings_pie_chart.png}
%   \caption{STT rating distribution}
%   \label{fig:stt_ratings}
%   \Description{Ratings of STTs in the actual-run experiment.}
% \end{figure}

% \begin{figure}[b]
%   \centering
%   \includegraphics[width=\linewidth]{full_run_ar_classifier_paper.png}
%   \caption{Time-saving offered with the blank classifier in actual-run}
%   \label{fig:ar_actual}
%   \Description{Performance of the blank classifier in actual run.}
% \end{figure}

% \begin{figure}[b]
%   \centering
%   \includegraphics[width=\linewidth]{full_run_moderation_time_saving_paper.png}
%   \caption{Time-saving offered with the gender classifier and location entity in actual-run}
%   \label{fig:trans_full}
%   \Description{Actual-run moderation time saving.}
% \end{figure}

% \begin{figure}[b]
%   \centering
%   \includegraphics[width=\linewidth]{full_run_jist_time_saving_paper.png}
%   \caption{Time-saving offered with the speech-to-text transcription in writing gist transcription in actual-run}
%   \label{fig:gist_actual}
%   \Description{Actual-run gist.}
% \end{figure}

% \begin{figure}[b]
%   \centering
%   \includegraphics[width=\linewidth]{full_run_full_time_saving_paper.png}
%   \caption{Time-saving offered with the speech-to-text transcription in writing full transcription in actual-run}
%   \label{fig:full_actual}
%   \Description{Actual-run full transcription time-saving.}
% \end{figure}

\subsection{Actual-run Experiment}
\begin{figure}[b]
     \centering
     \begin{subfigure}[b]{0.45\textwidth}
         \centering
         \includegraphics[width=\textwidth]{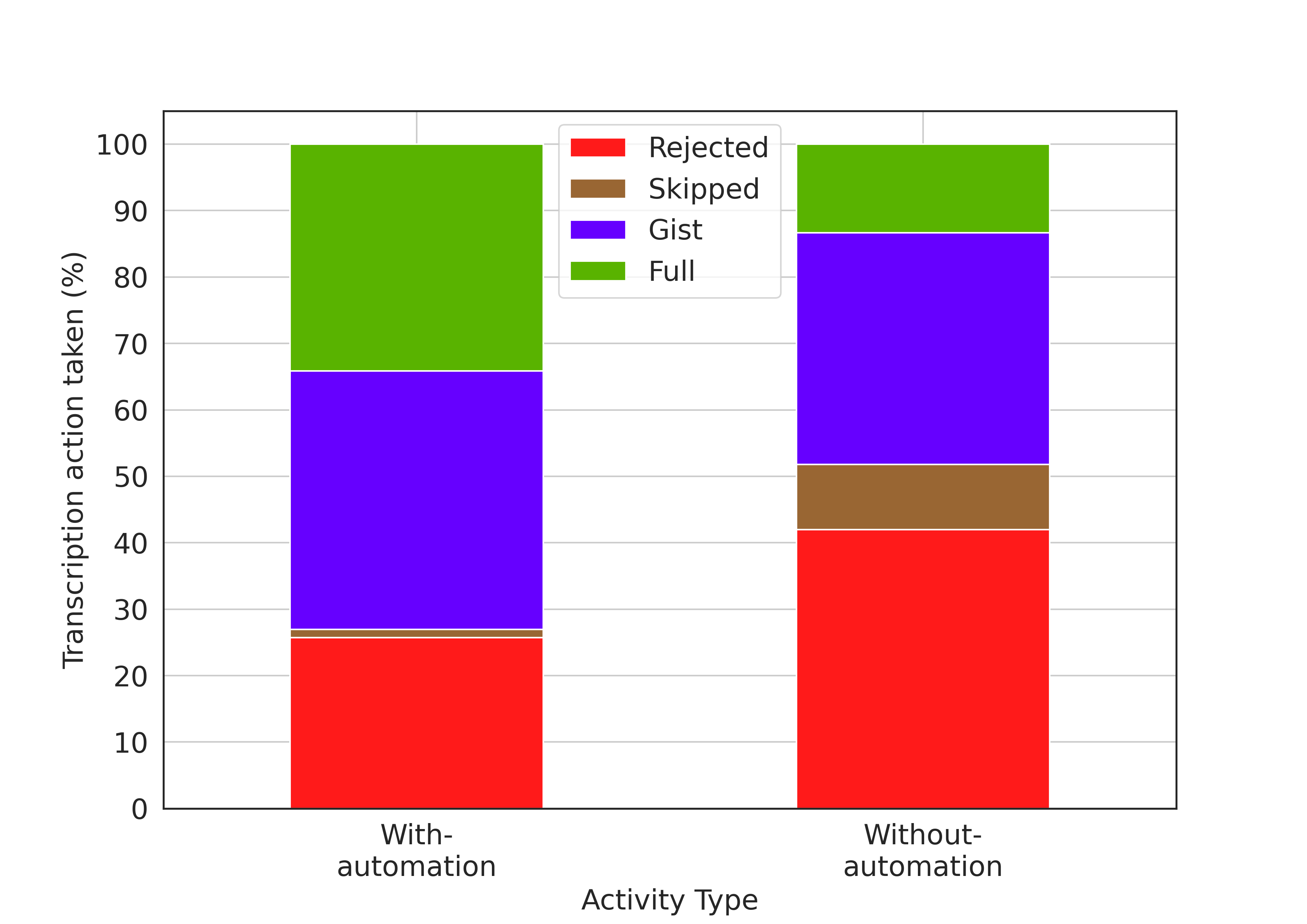}
         \caption{Difference between the action taken by moderators with and without automation}
         \label{fig:act_actual}
     \end{subfigure}
     \hfill
     \begin{subfigure}[b]{0.45\textwidth}
         \centering
         \includegraphics[width=\textwidth]{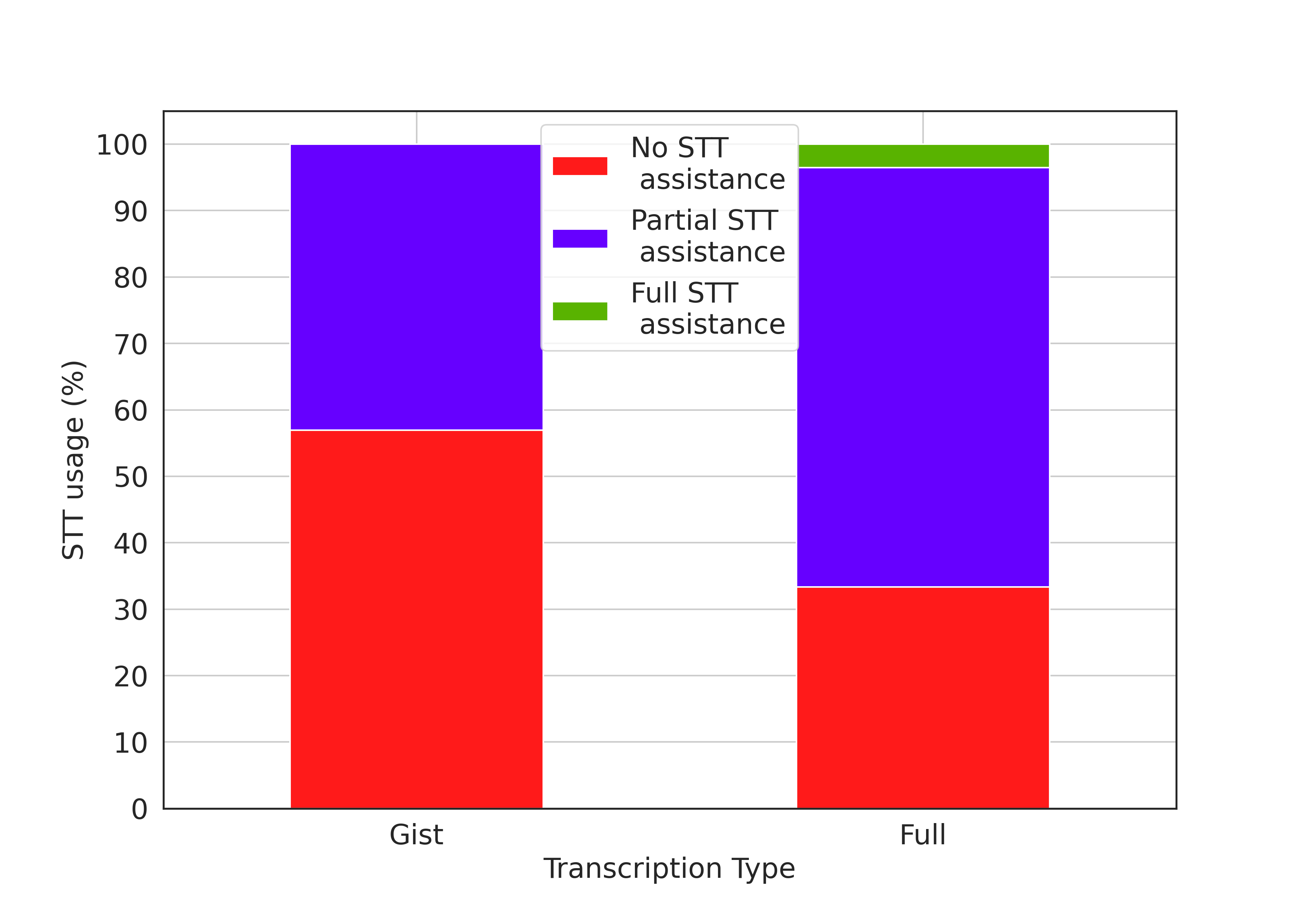}
         \caption{Use of STT to provide a gist or full transcription}
         \label{fig:stt_gist_full}
     \end{subfigure}
    %  \hfill
    %  \begin{subfigure}[b]{0.3\textwidth}
    %      \centering
    %      \includegraphics[width=\textwidth]{graph3}
    %      \caption{$y=5/x$}
    %      \label{fig:five over x}
    %  \end{subfigure}
        \caption{Actual-run experiment}
        \label{fig:act_run}
\end{figure}

\begin{table*}[t]
\caption{Item duration wise average time-saving}
\label{tab:agg_time_save}
\begin{tabular}{|c|c|c|c|c|c|}
\toprule
\multicolumn{1}{|p{1.5cm}|}{\centering Item \\ duration \\ bins}
%   & \multicolumn{1}{|p{1.2cm}|}{\centering \% items\\rejected \\ by blank \\  classifier} 
%   & \multicolumn{1}{|p{1.2cm}|}{\centering Avg. time\\saved \\ by blank \\ classifier (s)} 
   & \multicolumn{1}{|p{1.5cm}|}{\centering \% accepted\\items with \\ gist \\transcript}
   & \multicolumn{1}{|p{1.5cm}|}{\centering Avg. time\\saved in \\ writing \\gist\\transcript (s)}
   & \multicolumn{1}{|p{1.5cm}|}{\centering \% accepted\\items with \\ full \\transcript}
   & \multicolumn{1}{|p{1.5cm}|}{\centering Avg. time\\saved in \\ writing \\full\\transcript (s)}
%   & \multicolumn{1}{|p{1.2cm}|}{\centering Avg. time\\saved in \\ marking \\metadata (s)}
%   & \multicolumn{1}{|p{1.2cm}|}{\centering Avg. time\\saved \\ with blank\\classifier\\+metadata\\+STT}
   & \multicolumn{1}{|p{1.5cm}|}{\centering Avg. time saved in transcription (s)}\\
%   & \multicolumn{1}{|p{1.2cm}|}{\centering Total\\avg. time\\taken (s)}\\
%   \midrule
%   0-10 & 100 & 16.43 & 33.33 & 0.00 & 66.67 & 0.00 & 0.00 & 16.43 & 27.10\\
  \midrule
  10-20 & 75.00 & -1.25 & 25.00 & -0.57 & -1.08\\
  \midrule
  20-40 & 60.00 & 3.13 & 28.00 & -25.06 & -5.14 \\
  \midrule
  40-60 & 75.00 & 11.52 & 20.00 & 39.16 & 16.47 \\
  \midrule
  60-100 & 59.62 & 9.29 & 30.77 & 102.99 & 37.50 \\
  \midrule
  100 or more & 56.10 & 10.61 & 36.59 & 208.00 & 82.90 \\
%   \midrule
% \multirow{2}{*}{Location} & State    & \multirow{2}{*}{28.71} & 51.81 & 81.48 \\
%                           & District &                        & 51.15 & 65.34 \\
% \midrule
% \multirow{3}{*}{\begin{tabular}[c]{@{}c@{}}Date of\\  birth\end{tabular}} &
%   Date &
%   \multirow{3}{*}{19.65} &
%   88.88 &
%   98.88 \\
%                           & Month    &                        & 88.88 & 98.88 \\
%                           & Year     &                        & 84.44 & 98.88 \\
% \midrule
% Number                    &          & 26.47                  & 93.33 & 94.66 \\
\bottomrule
\end{tabular}
\end{table*}
As we have seen so far, the use of automation does lead to time-savings, and we wanted to see how the moderators choose to use this time. We therefore ran an additional experiment with and without automation, as before, but this time we removed the restriction of moderating for 6 hours only. We wanted to create as natural an environment as possible so that the moderators could freely allocate their time to tasks they believed to be important. Additionally, to understand the reasons behind the actions taken by them especially on utilizing the STT transcripts, we asked them to also note the actions they took, as follows: 

\begin{itemize}
    \item Skipped transcription: Moderators skipped writing a transcript for the item
    \item Transcription type: Whether the item was given a gist or a full transcription
    \item No STT assistance: The STT had too many errors, or the information was not useful, to utilize it for a gist or full transcription
    \item Partial STT assistance: The moderators copied parts of the STT transcript and made edits
    \item Full STT assistance: The moderators copied the STT transcript as such
%- Transcription Type: Moderators chose to write a full transcription or a gist transcription
%- Total ML: Few erroneous words and the STT was allowed to go as such
%- Partial ML: Moderators copied the ML transcript but made some corrections/additions on the same
%- No-ML full: Too many erroneous words in an item which ideally required a detailed transcript so complete transcription had to be written by Moderators
%- No-ML gist: Too many erroneous words and the complete third-person gist is written for that item by the Moderators
\end{itemize}

%Our motivation of conducting this experiment was influenced from our findings of the previous experiment in which we found that the Moderators were not able to take the complete benefit of the STT. In this experiment, we dug deeper to understand how Moderators use automation when the moderation policies are left up to their own discretion. Hence, along with the time annotations, accept/reject decision and reason for rejection, we asked the Moderators to mark two more field in the excel data sheet which were \emph{STT Rating(1-5)} and \emph{Action taken}. The STT rating defined the quality of the STT on a scale of 1 to 5 and 0 if not present. The action field could be any one of the 5 fields:

%This actual-run experiment was also conducted with and without automation but here the restriction of marking the items only for 6 hours each day been eased because we were studying the activity of Moderators when no restriction had been imposed on them. The "with-automation" part of the experiment could have any one of the 5 actions while the without automation part could only have any one of the last three actions taken by the Moderators.

In this experiment, we saw a total of 422 items, out of which 167 (of which 124 were accepted for publication) were studied during the with-automation phase, and 255 (of which 148 were accepted for publication) were studied during the without-automation phase.

% \begin{figure}[h]
%   \centering
%   \includegraphics[width=8cm]{transcription_type_dist_paper.png}
%   \caption{Difference between the action taken by the Moderators with and without automation}
%   \label{fig:act_actual}
%   \Description{Actual-run action distribution}
% \end{figure}

% \begin{figure}[b]
%   \centering
%   \includegraphics[width=8cm]{ml_trans_action_dist_paper.png}
%   \caption{Use of STT to provide a gist or a full transcription}
%   \label{fig:stt_gist_full}
%   \Description{Actual-run STT use in gist and full transcription.}
% \end{figure}

Figure \ref{fig:act_actual} shows the differences in the actions taken by the moderators with and without automation. The most marked difference is that the time-saving during automation seems to have been used to provide more full transcripts and less of skipped items than without automation. We found this to be interesting that moderators felt that many more items deserved to be transcribed than what they were able to do without automation. This was confirmed during feedback sessions with the moderators, where many of them reported that they often faced a time crunch and were not able to give due attention to many useful items. In Figure \ref{fig:stt_gist_full}, we further analyzed the use of STT transcripts to provide a gist and full transcription. While \emph{Full STT assistance} for full transcripts was rare, \emph{Partial STT assistance} was taken actively by the moderators to write full transcripts, and also to write gist transcripts to some extent. In about 30\% of the cases for full transcripts, \emph{No STT assistance} was taken at all due to poor STT transcript quality. Overall, we found that the AI-based tools were accepted by the moderators and they did have an impact in terms of shifting the emphasis on various tasks done by the moderators.

\subsection{Aggregate Time-cost Savings}
We use the results of the actual run experiment to calculate the aggregate time saving. Table \ref{tab:agg_time_save} shows the percentage distribution of accepted items across the different bins, the average time-saving per item attributed to automated gist and full transcription respectively.
\begin{comment}
\textcolor{green}{Table \ref{tab:agg_time_save} shows for different bins, the percentage of items that were rejected for being blank, and the time-saving made per item by the blank classifier; and similarly for accepted items, the percentage distribution of accepted items across the different bins, the average time-saving per item attributed to automated metadata annotation, and for gist and full transcription respectively.}
\end{comment}
Through this aggregate analysis, with the assumption that the arrival pattern of items remains invariant across the audio duration bins, blank and acceptable items in each bin, and those deserving gist or full transcription in each bin, we find that automation is able to provide a time-saving of approximately 40\% per item (an approximate time-saving of 54 seconds per item out of the average time of 134 seconds taken to moderate an item) with the help of automation.
\begin{comment}
\textcolor{red}{where the average time taken to moderate an accepted item was 134.45 seconds and we were able to save 54.13 seconds on an average, with the help of automation}
\end{comment}
This saving, as we saw, helps the moderators provide additional transcriptions, or can be used towards other useful tasks such as calling the users to seek feedback regarding their participation on the voice forums, or provide guidance to the users to record better audio messages. 

Converting to cost terms, we use an average moderator salary as INR 20,000 per month, with 48 hours of work per week, 15 items moderated per hour, and an additional cost overhead of 30\% to account for office space, utilities, etc. This leads to an average per-item moderation cost of INR 8.3. A 40\% time-saving results in an improved per-item moderation cost of INR 4.98. However, to this we need to add the cost of the Google ASR APIs used to obtain the STT transcripts. The APIs cost INR 0.29 for a 15 seconds audio, and aggregating over the same distribution of audio duration gives an average STT transcript cost of INR 1.45 per item. We inflate this by applying an overhead cost of 30\% for additional technology management and other expenses like the use of cloud computing infrastructure. Adding it to the per-item moderation cost with automation, this gives us a final cost-saving of almost 17\% with automation. The automation therefore provides us with both cost-saving as well as time-saving. The time-saving is more significant and can be used towards tasks which otherwise were getting ignored or sidelined in favour of various necessary routine tasks.

\section{Discussion}

\subsection{Other Opportunities}
The benefits of AI-based automation both in terms of time and cost saving encourages us to find other opportunities for AI-based tools as well. A few such options include: 

\begin{itemize}
    \item \emph{Detection of noisy items}: Just like the blank-items classifier, we are building a classifier to detect noisy items which can be automatically rejected. We however found that the same audio features we used for the blank and gender classifiers, are not adequate for the noisy classifier since many audios are noisy due to background human voice and a TSNE plot of the features showed that there is no clear separation between noisy and accepted items. Therefore, we developed a CNN (convolutional neural network) based classifier which used the mel scaled audio spectrogram \cite{brian_mcfee_2021_4792298} as input to the classifier and achieved 98.2\% accuracy and a false negative rate of 3.6\%. We are now in the process of deploying this model in production and are also trying to improve upon the model accuracy by augmenting our internal dataset with an external dataset containing 50 environment sounds \cite{piczak2015dataset}.
    
    \item \emph{Tag annotation}: Tags related to broad topics such health, agriculture, education, governance, etc. form a significant part of metadata annotation done by the moderators. We are building multi-label tag classifiers based on the STT transcripts, to pre-assign tags to the audio items. Simple keyword based classifiers using word vectors or TFIDF weighted bag-of-words approaches are not able to give a very good performance, and we are looking at other approaches that can utilize underlying hierarchical structures in the tag assignments as well. 
    \item \emph{User interfaces for editing transcripts}: The STT transcripts obtained through the Google ASR APIs also provide confidence scores for each word in the transcript. When the STT transcripts are presented to the moderators, highlighting words predicted with low-confidence in the STT transcripts may make it easier for the moderators to spot errors in the transcripts and selectively fix these errors. If such improvements in the user interface make it easier for the moderators to utilize the STT transcripts then it is likely to lead to further time savings.
    \item \emph{Continuous model improvements}: We evaluated the performance of our deployed machine learning models to observe the change in accuracy of such systems from a test environment to a production environment. We observed a performance dip of 3-4\% among both the blank classifier as well as the gender classifier. %with the blank classifier achieving a 94\% production accuracy with a false negative rate of 7\% and the gender classifier achieving 88\% production accuracy with an F1 score of 0.93.
    We further examined if any gender bias had crept into the false negatives predicted by the blank classifier but found that the model performance was unbiased and false negative rate was equal among both the genders.
    % \begin{comment}
    % In order to improve upon these negative predictions by our machine learning models in production,
    % \end{comment}
    To improve production accuracy, we are using a human-in-the-loop approach \cite{wang2021putting} to build an automated pipeline which incorporates the feedback of the moderators on a monthly basis and fine-tunes the trained models through hard negative mining. Such an approach should help in a continuous model improvement and prevent the models from drifting in the production environment. 
    \item \emph{Identification of erroneous predictions}: Our interaction with the Gram Vaani moderators revealed that they would benefit from seeing a confidence score against the prediction results of the various classifiers, since it would give them hints on which contributions to examine more carefully than others. This is in line with HCI best practices suggested for the incorporation of AI-based automation in risk assessment tool \cite{de2020case} and should help in identification and correction of erroneous classifier predictions.
    %in our current version of the moderation interface, it was difficult for the moderators to identify incorrect predictions made by the machine learning classifiers as the moderators had to listen to the audio recordings and cross-check if the classifier output was in-fact correct. A better design approach for incorporating AI predictions in content moderation systems would be to display the confidence score of predictions by the classifiers in the moderation interface along with the action of automatic audio rejection, metadata annotation and STT transcription. Moreover, this annotation should not be done at all when the classifier confidence is lesser than a certain threshold value. Such a design has helped humans override algorithmic prediction in risk assessment tools \cite{de2020case} and should similarly help in identification and correction of the incorrect predictions by the machine learning classifiers in the automatic content moderation systems.
    \item \emph{Distributed moderation}: Centralized moderation stands the risk of making incorrect editorial decisions due to a lack of knowledge of the local context. This may include decisions on ranking and exposure given to different content, recognition of misinformation, acceptance/rejection of contentious contributions, etc. We have not studied such editorial decision making processes as yet among the content moderators, and the impact of AI-based automation on these processes. Since the Mobile Vaani platforms are supported by a team of offline community volunteers who are well placed to take more informed editorial decisions, as part of future work, we plan to also study editorial decision making in centralized and distributed moderation settings, with and without AI-based automation tools such as the ones described in this paper. Several interesting questions will be worth investigating, of how to arrive at a consensus in the editorial decisions taken by a group of volunteers, how do these compare with decisions made by the central team of moderators, time-saving among volunteers and moderators through the use of automation tools, and alternate activities for which time was created such as to guide community members to make better contributions or to connect more deeply and frequently with the community members. We believe that AI-based tools can also be helpful to identify these community members who would benefit from guidance by the volunteers and reduce user dropout.
    
    %In future, we would also like to evaluate the effect of automation in distributed moderation systems. Most likely, such systems will open up new questions on comparison of editorial decisions between the community volunteers and the moderators, in terms of recognition of misinformation, higher ranking of locally relevant content in the community and creation of debate during decision mismatch. It will hence become important to evaluate the extent to which automation helps volunteers connect better with their communities which could be in easier identification of users to whom guidance calls should be given in priority and in identification of users who could be on the fence of dropping out of the use of platform and need to be encouraged or interviewed to understand what is missing from the platform and how to make it more exciting for them to use.
    
    % and to help address this increase in demand, we quickly deployed the initial versions of our AI models to support the in-house Moderators.  }
    
\end{itemize}

\subsection{Use of Automated Tools During COVID-19}

\begin{figure}[b]
  \centering
  \includegraphics[width=\linewidth]{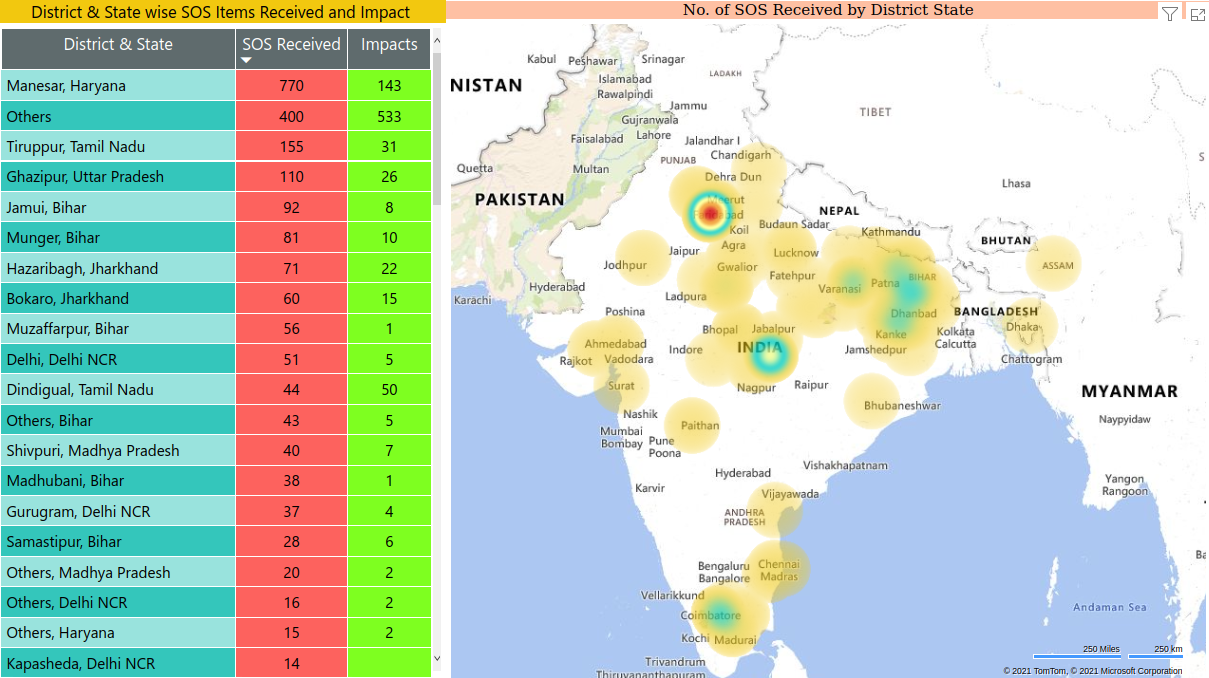}
  \caption{Location-wise grievance recordings as visible on Gram Vaani's live dashboard.}
  \label{fig:dashboard}
  \Description{Location grievances data statistics}
\end{figure}

The Gram Vaani voice-forums were heavily used during the COVID-19 lockdown in India, which was announced towards the end of March 2020. An overnight order for suspension of factories and transport caused wide distress among migrant workers who were stranded in cities, rural populations who faced food shortages due to broken supply chains, and significant income loss due to stoppage of agricultural and daily wage work \cite{indiaforum}. People called into the IVR platforms to report about problems they were facing, described their experiences, and asked for help. Over a million people used the platforms during the first few months of the COVID-19 lockdown, and contributed over 20,000 voice recordings \cite{100daystudy}. The content moderators were naturally overwhelmed with this sudden spike and we conducted a rapid deployment of an initial version of some of the moderation automation tools we have presented in this paper. We first bifurcated content contributions in the IVR itself by asking people whether they wanted to seek emergency help or to contribute a voice report. The voice recordings of those asking for help was transcribed through the ASR APIs and the location entity extraction module was used to identify the place from which people were calling. In some geographies, these recordings were immediately passed on to the Gram Vaani volunteers from that location to provide assistance. In the Delhi National Capital Region where migrant workers were stranded and rendered out of cash and food, the volunteers were able to quickly connect them with humanitarian organizations working on the ground and managed to help 1,800 workers with food kits and cash transfers \cite{lockdownchronicals}. Several other organizations also requested for a rapid deployment of the IVR and were able to assist several thousand workers with transportation to their villages by registering their demand and using it to persuade governments to arrange trains for their safe transport \cite{covid19response}. The moderators appreciated the benefit of this automated segregation and delegation, which reduced the workload on them: 

``\emph{During Coronavirus, the Saajha Manch project received a huge number of news recordings and during that time ML helped us a lot. In a day, our team of 4 moderators were able to moderate 400 items each}''.

We also built a simple word frequency based categorization to understand the nature of the problems being reported by the people, such as: \emph{out of food, stuck in city, health emergency, out of cash, not received government cash transfers, bank not accessible, black marketing and price rise of food, gas relief not received, social distancing not being followed, issues at isolation centers, agriculture and other livelihood related issues}, etc. We then clubbed this category information with the output from the location entity extractor to map the issues on our own live dashboard (as shown in figure \ref{fig:dashboard}) and aggregators like \emph{MapMyIndia} \cite{MMI}. These dashboards helped portray the extent and intensity of the distress experienced by the people and were used in PILs (Public Interest Litigations) filed by activists to draw the government's attention to the huge distress induced by the sudden and stringent lockdown.

\section{Conclusions}

Overall, we found that the AI-based tools we built were able to give modest cost-saving of 17\% but substantial time-saving of 40\% for content moderation in voice-based forums. These tools cannot replace human moderators but can augment their work. This nuance is different from oft-heard analysis of impending unemployment due to AI-based automation; our experience rather suggests that AI-based tools can improve the productivity of human workers rather than replace them, and the time thus saved can be dedicated for other important tasks that may have been ignored so far. Insights from the detailed instrumentation exercise conducted by us can be useful for other researchers and practitioners working with voice-based tools to automate their operations. 

\begin{acks}
We would like to express our immense gratitude to the entire moderation team for helping us conduct this study and providing us with daily feedback and results for our time-cost analysis. We would also like to thank the entire Gram Vaani team who provided us with constant support in conducting this project. We would also like to thank the ACT4D lab at IIT Delhi for providing us with the initial compute infrastructure. Finally, we would like to thank the Bill \& Melinda Gates foundation for funding this project.
\end{acks}

\bibliographystyle{ACM-Reference-Format}
\bibliography{sample-base}

%%
%% If your work has an appendix, this is the place to put it.
\appendix
\end{document}